\documentclass{pasj00}

\def\tsutsui{$E_{\rm p}$--$T_{\rm L}$--$L_{\rm p}$ }

\SetRunningHead{R. Tsutsui and T. Shigeyama}{Scaling law in long gamma-ray bursts}

\begin{document}

\title{Universal scaling law in long gamma-ray bursts}
\author{Ryo \textsc{Tsutsui},\altaffilmark{1}\email{tsutsui@resceu.s.u-tokyo.ac.jp} \& Toshikazu \textsc{Shigeyama}, \altaffilmark{1}}
\altaffiltext{1}{Research Center for the Early Universe, School of Science, University of Tokyo, Bunkyo-ku, Tokyo 113-0033, Japan}


%


\maketitle

\begin{abstract}
Overwhelming diversity of long gamma-ray bursts (LGRBs), discovered after the launch of  {\it Swift} satellite
, is a major obstacle to LGRB studies. 
Recently, it is shown that the prompt emission of LGRBs is classified into three subclasses: Type I, Type II LGRBs populating separate fundamental planes in a 3D space defined by the peak luminosity, the duration, and the spectral peak energy, and outliers not belonging to either of the planes. 
Here we show that Type I LGRBs (LGRBs I) exhibit different shapes of light curves from Type II LGRBs (LGRBs II). Furthermore, we demonstrate that this classification has uncovered a new scaling law in the light curves of LGRBs II spanning 8 orders of magnitude from the prompt to late X-ray afterglow emission. The scaled light curve has four distinct phases. The first phase has a characteristic time scale while the subsequent three phases exhibit  power law behaviors with different exponents.
We discuss its possible interpretation in terms of the emission from an optically thick fireball propagating in the cricumstellar matter at relativistic speeds and argue that the observed four phases correspond to its hydrodynamical phases. Our classification scheme can pin down the intrinsic luminosities of LGRBs II through the scaling law from a sample of polymorphic GRBs. Further refinement of this scheme and scaling law will make a subclass of LGRBs a standard candle as reliable and accurate as Type Ia supernovae in the more distant universe than supernovae can reach. 
\end{abstract}

\section{Introduction}
Gamma-ray burst (GRB) is the most energetic explosion in the universe.  A GRB of which the prompt emission  lasts longer than 2 seconds is classified into a long GRB (LGRB). 
The prompt gamma-ray emission is followed by afterglow emission at longer wave-lengths. 
LGRBs are thought to come from the death of massive stars.
The discovery of supernova 2003dh associated with a long GRB 030329\citep{Stanek:2003} was thought to conform such a hypothesis. 
However, observational success after the launch of {\it Swift} satellite\citep{Gehrels:2004} casted doubt on the bimodal classification based on the duration of the prompt emission. 
In spite of its extremely long duration ($\sim 102$ sec), GRB 060614\citep{Gehrels:2006} could result from  a neutron star merger because of no associated supernova. 
Even if associated supernovae are observed, many of them, e.g. GRBs 980425 and 060218, are far dimmer than ordinary GRBs. 
In addition to these mysteries, diversity of the early X-ray afterglow discovered with the X-ray Telescope (XRT) on board {\it Swift} satellite is another major obstacle in the field of GRB study.
Though all of these observations indicate a limitation of short-long dichotomy of gamma-ray bursts,  until now there has been no successful classification scheme to take into account all of these observational properties consistently. 

Well known correlations between observed quantities in prompt emission such as the spectral peak energy ($E_{\rm p}$) -- isotropic energy ($E_{\rm iso}$),
 $E_{\rm p}$ -- peak luminosity ($L_{\rm p}$), or  spectral lag ($t_{\rm lag}$) -- peak luminosity ($L_{\rm p}$) 
correlations\citep{Amati:2002,Yonetoku:2004,Norris:2000} are sometimes used as discriminators of GRBs\citep{Lu:2010,Gehrels:2006}. 
These correlations that divide GRBs into short GRBs (SGRBs), LGRBs, and some low-luminosity GRBs, 
have left a large dispersion in the correlation of each class.
Though the dispersion had been implicitly assumed to follow a single gaussian distribution in many studies before {\it Swift} and {\it Fermi}, 
recent observations of no associated supernova with an apparent LGRB and diversity in X-ray afterglow might have indicated that this is not the case.   
Recent observations, by {\it Swift} and {\it Fermi} etc, provide so plenty of data that this implicit assumption can be tested.

Thanks to thorough studies to reduce systematic errors in $E_{\rm p}$, $L_{\rm p}$, $E_{\rm iso}$ and the duration of GRBs\citep{Kaneko:2006,Yonetoku:2010,Kocevski:2012}, 
criteria for the gold sample of GRBs were established as follows\citep{Tsutsui:2011,Tsutsui:2012a}:  
\begin{itemize}
\item[1.] Time integrated spectra must be fitted with the Band model to reduce systematic errors due to different choices of spectral models.
\item[2.] Light curves must have a time resolution of 64 msec  to derive the peak luminosities  with the same time resolution in the GRB rest frame by re-binning  the fluxes.
\item[3.] Peak photon counts must be 10 times larger than the background fluctuation for secure estimations of the duration and $E_{\rm iso}$.
\end{itemize}

The gold sample compiled by \citet{Tsutsui:2012a} had the smallest systematic error and made it possible to find out 
a generalized correlation between $E_{\rm p}$, $T_{\rm L}$ ($\equiv E_{\rm iso}/L_{\rm p}$) and $L_{\rm p}$.
Furthermore, the correlation succeeded in discriminating, at least, three subclasses in LGRBs\citep{Tsutsui:2011,Tsutsui:2012a}  : 
Type I, Type II, and outliers. 
This new classification was verified from the temporal properties of the prompt emission\citep{Tsutsui:2012a}. 
Figure 1a shows the \tsutsui diagram of the gold sample with the classification in \citet{Tsutsui:2012a}.
Blue squares indicate LGRBs I populating a fundamental plane plotted with the solid line,
 red circles LGRBs II on another slightly misaligned fundamental plane, 
 and black squares outliers.
LGRBs I and II are also discriminated by the shape of their prompt emission light curves.
Figure 1b shows typical light curves for LGRBs I  (990506) and II (081222).
As exemplified in this figure, LGRBs I have many spiky pulses with sharp declines, while LGRBs II have consecutive pulses with short intervals followed by long tails. 
To clarify the difference, we make a plot of normalized cumulative counts as functions of  time  normalized with $T_\mathrm{98}$ (the period during which 98\% of the total flux is received) for 17 LGRBs I and 11 LGRBs II in Figure 1c.
LGRBs II (red curves)  deviate from the constant count rate (black solid line) further than  LGRBs I(blue curves) (See \citet{Tsutsui:2012a} for more details).

In this letter, we compile combined BAT - XRT light curves for each subclass in the gold sample and discuss differences in the shapes of light curves of GRBs belonging to different subclasses in section 2. We analyze light curves of LGRBs II in particular to seek universal features covering all the phases from the prompt to late X-ray afterglow emission in section 3. Finally, we present a possible interpretation using a fireball model in section 4.

\begin{figure*}[htbp]
\begin{center}
\begin{tabular}{c}
\begin{minipage}{0.50\hsize}
\begin{center}
\includegraphics[clip,width=9cm]{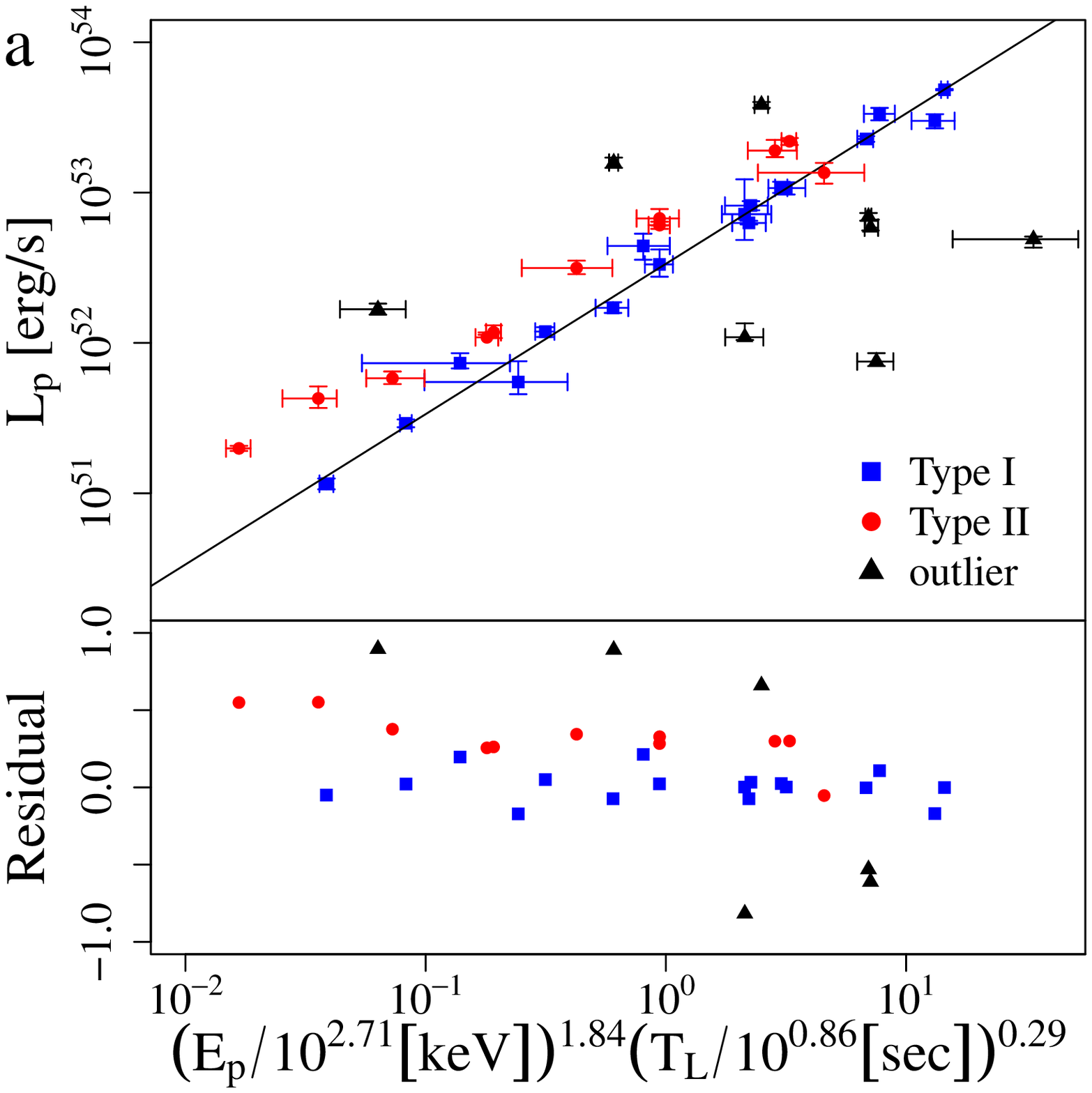}
\end{center}
\end{minipage}
\begin{minipage}{0.50\hsize}
\begin{center}
\includegraphics[clip,width=9cm]{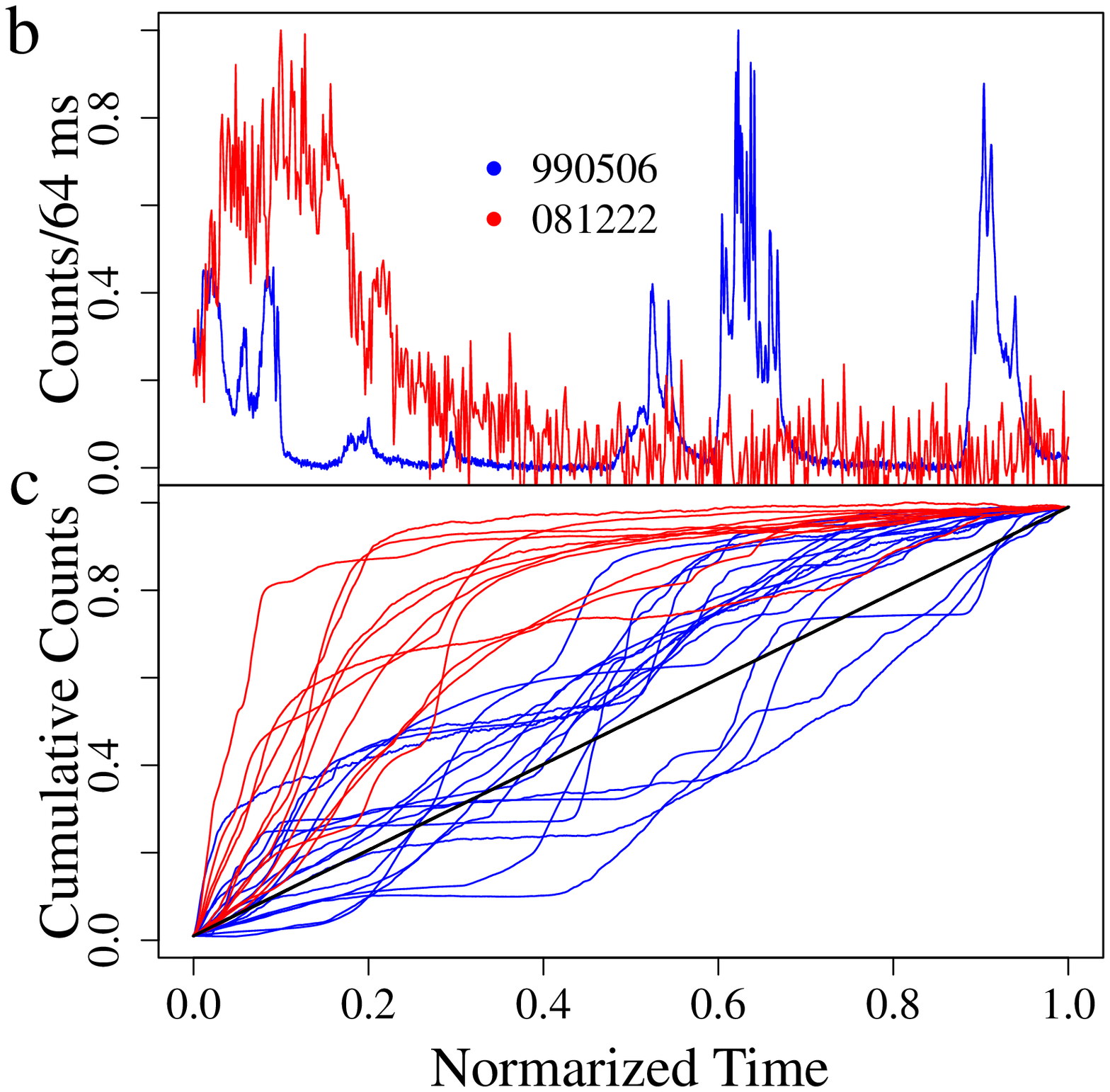}
 \end{center}
\end{minipage}
\end{tabular}
\end{center}
\caption{{\bf Classification scheme of prompt emission of long GRB.} 
{\bf a}, The \tsutsui diagram for 36 long GRBs in the gold sample in \citet{Tsutsui:2012a} with the residual plot in the bottom panel. 
Three subclasses based on the distribution in this plot are indicated by different colors: Type I (blue), Type II (red) and outliers (black). The same color is used for the same subclass in the other panels.
The solid black line indicates the fundamental plane for Type I long GRBs.
{\bf b}, Examples of normalized light curves. 
{\bf c}, Normalized cumulative light curves. The black solid line indicates a constant count rate.
}
\label{allGRB}
\end{figure*}

\begin{table*}

\caption{Spectroscopic redshifts, class, $T_{\rm L}$, $L_{\rm p,51}(\equiv L_{\rm p}/(10^{51}~$erg/s)) of the gold sample observed by {\it Swift}/XRT from \citet{Tsutsui:2012a}.
To minimize the dispersion in the \tsutsui diagram, $L_{\rm p}$ were re-binned  by 2.752 sec in the GRB rest frame.
}
\begin{center}
\begin{tabular}{cccccccccc}
\hline \hline
GRB & z &class& $T_{\rm L}$ & $L_{\rm p,51} $&GRB & z &class& $T_{\rm L}$ & $L_{\rm p,51} $\\ 
	&		&	& [sec] 	& [erg/sec$^{-1}$]&	&		&	& [sec] 	& [erg/sec$^{-1}$] \\ \hline
050401	&	2.9	&Type I	&	7.00	&	7.34	&	090424	&	0.544	&Type II	&	3.97	&	4.27	\\
050525	&	0.606	&Type I	&	7.95	&	71.80	&	090618	&	0.54	&Type I	&	21.7	&	69.05	\\
050603	&	2.821	&Type II	&	3.51	&	227.31	&	090902B	&	1.822	&Type I	&	6.37	&	5.82	\\
061007	&	1.261	&Type I	&	9.13	&	81.96	&	090926A	&	2.106	&outlier	&	5.00	&	334.17	\\
070125	&	1.547	&Type I	&	8.60	&	33.29	&	091003	&	0.897	&outlier	&	8.40	&	62.85	\\
071003	&	1.604	&outlier	&	8.35	&	17.10	&	091127	&	0.49	&Type II	&	8.45	&	67.297	\\
071010B	&	0.947	&Type II	&	5.50	&	219.39	&	091208B	&	1.063	&Type I	&	4.41	&	299.80	\\
080319B	&	0.937	&outlier	&	19.26	&	1.16	&	100414A	&	1.368	&outlier	&	13.0	&	7.52	\\
080413B	&	1.1	&Type II	&	3.18	&	44.17	&	100906A	&	1.727	&Type II	&	9.13	&	11.773	\\
080721	&	2.602	&Type I	&	3.59	&	2.92	&	110213A	&	1.46	&outlier	&	4.80	&	11.9	\\
081121	&	2.512	&Type I	&	3.95	&	135.53	&	110422A	&	1.77	&outlier	&	4.54	&	483.4	\\
081222	&	2.77	&Type II	&	3.14	&	107.56	&	110503A	&	1.613	&Type II	&	2.96	&	384.026	\\
090323	&	3.57	&Type I	&	12.84	&	107.14	&	110715A	&	0.82	&Type II	&	4.42	&	10.905	\\
090328	&	0.736	&outlier	&	22.61	&	48.96	&	110731A	&	2.83	&Type II	&	2.39	&	2.0	\\ \hline
\end{tabular}
\end{center}
\label{tab1}
\end{table*}%

\section{X-ray afterglow}
The classification scheme was developed with  information only of the prompt emission. Nonetheless, it probably has a link with the afterglows.
Here we investigate the temporal profile of combined prompt - X-ray afterglow emission for each subclass of LGRBs.  
To do this,  we obtain combined BAT-XRT  flux light curves in the energy range of 15-50 keV for our gold sample from the Swift Burst Analyser website\footnote{http://www.swift.ac.uk/burst\_analyser/}\citep{Evans:2010}.
We choose BAT data  binned with the signal-to-noise ratio of at least 5.
{\it Swift}/XRT observed 28 out of 36  LGRBs in the gold sample.  
The present sample with {\it Swift} observations is composed of 10 LGRBs I, 10 LGRBs II, and 8 outliers.
In Figure 2, we show combined BAT-XRT light curves of all the 28 LGRBs.
The reddish  points indicate the light curves of LGRBs II, 
the blueish points LGRBs I, and the blackish points outliers. 
Roughly speaking,  the data for the first 100 sec are taken with the BAT and the rest  with the XRT.
As the figure shows,  LGRBs II have three breaks in their light curves in common, while LGRBs I and outliers exhibit much complicated behaviors.
We will discuss the early X-ray afterglow emission of LGRBs I and outliers in separate papers. Here we concentrate on the analysis of LGRBs II, which is much simpler than the other types.

\begin{figure*}[htbp]
\begin{center}
\begin{tabular}{c}
\begin{minipage}{0.50\hsize}
\begin{center}
\includegraphics[clip,width=9cm]{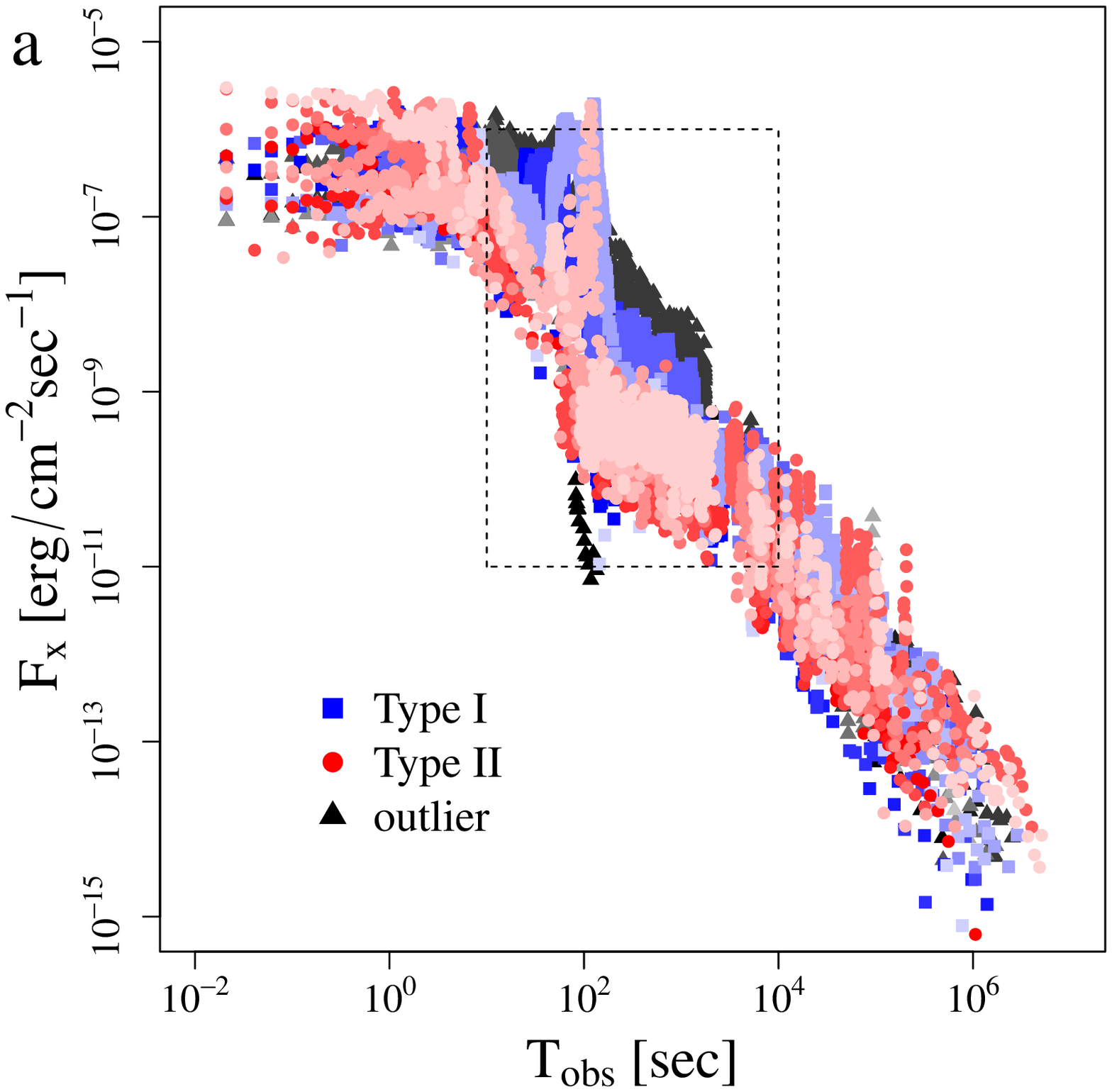}
\end{center}
\end{minipage}
\begin{minipage}{0.50\hsize}
\begin{center}
\includegraphics[clip,width=9cm]{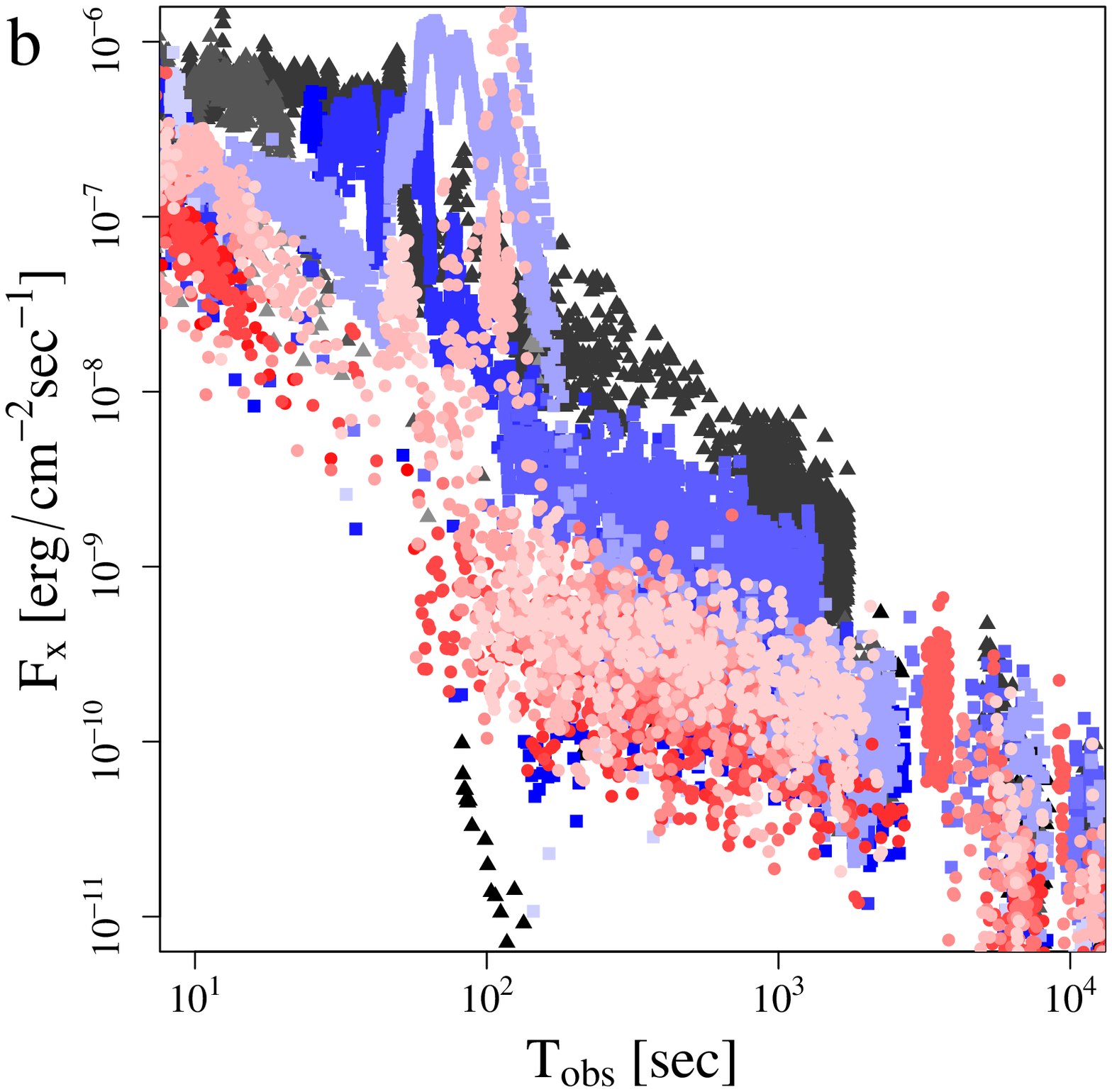}
 \end{center}
\end{minipage}
\end{tabular}
\end{center}
\caption{{\bf Combined BAT-XRT light curves for each subclass of the gold sample. }
 {\bf a}, Overall shape of the light curves of GRBs belonging to each subclass
 (blueish, Type I: redish, Type II: blackish, outliers).
 Individual events in the same subclass are distinguished by gradation.
 {\bf b},  Close-up of the region enclosed by a rectangle in the left panel. 
 
 }
\label{allGRB}
\end{figure*}

\section{Scaling law}
To seek a universal scaling relation in the light curves of LGRBs II, we normalize 15-50 keV luminosity $L_{\rm X}$ with $L_{\rm p}$ and the time  $T=T_{\rm obs}/(1+z)$ measured from the BAT trigger in the GRB rest frame with $T_{\rm L}$, where $T_{\rm obs}$ denotes the observed time, $z$ the redshift of the GRB host galaxy.
In Table 1, we summarize redshifts, classes, $T_{\rm L}$, and $L_{\rm p}$ for the gold sample observed by {\it Swift}/XRT.
In Figure 3a, we make a plot of normalized luminosity $L_{\rm X}/L_{\rm p}$ versus normalized time $T/T_{\rm L}$ for LGRBs II. We find that two of them, showed in the inset, 
have strong X-ray flares and exclude them from the following analysis. These two events also seem to follow the same scaling law for the others, except for X-ray flares. This suggests that the existence of X-ray flares does not affect the rest of the afterglow emission. 

We found that the lack  of data between  the first and second breaks for some GRBs makes it difficult to fit a function to the light curve of each LGRB II.
Instead of fitting to data of each  LGRB II,  we stack data of  the  8 LGRBs II thus selected to investigate the average properties and fit a broken power law model with three breaks. 
We use the least square method without any weighting factor, because the dispersion of the data around the best fit model is larger than error bars.
Therefore we minimize the following least square merit function, 
\begin{equation}
\label{ls}
S =\sum_{i}^{N} \left\{\log \left(\frac{L_{{\rm X},i}}{L_{\rm p}}\right) - \log \left(\frac{L_{\rm X,model}\left(\frac{T_i}{T_{\rm L}}\right)}{L_{\rm p}}\right)\right\}^2, 
\end{equation} 
and define the standard deviation as $\sigma \equiv \sqrt {S_{\rm min}/(N-7)}$.
This model has  7 parameters : the normalized luminosity $L_{\rm norm}$ in the prompt (first) phase,  the power law exponent for the $i$-th phase, $\alpha_i\,(i=2,\,3,\,4)$,  the first break time, $t_{\rm b1}$, second break time, $t_{\rm b2}$ and the third break time $t_{\rm b3}$. 
As a result, we obtain the best fit model, 
\begin{equation}
\label{eq}
\frac{L_{\rm X, model}(t)}{L_{\rm p}} =
 \left \{ \begin{array}{ll}
 C_1 & (\frac{T}{T_{\rm L}}<0.48)\\
 C_1  \left(\frac{T}{0.48 T_{\rm L}}\right)^{-1.99}    & (0.48 < \frac{T}{T_{\rm L}}<15.8) \\
   C_1 C_2 \left(\frac{T}{15.8T_{\rm L}}\right)^{-0.49} & (15.8< \frac{T}{T_{\rm L}}<273.6)\\
    C_1 C_2 C_3\left(\frac{T}{273.6T_{\rm L}}\right)^{-1.31} & (273.6< \frac{T}{T_{\rm L}})
  \end{array} \right. ,
\end{equation}
where $C_1=0.173$, $C_2= (15.8/0.48)^{-1.99}$ , and $C_3=(273.6/15.8)^{-0.49}$.
The best fit values ($1$-$\sigma$ errors)  for these parameters are $L_{\rm norm}=0.173 (0.0058)$, $\alpha_2=-1.99 (0.026)$, $\alpha_3=-0.49 (0.038)$, $\alpha_4=-1.31 (0.018)$, $t_{\rm b1}=0.48 (0.013)$, $t_{\rm b2}=15.83 (0.92)$, and $t_{\rm b3}=273.6 (27.6)$ with standard deviation $\sigma = 0.84$.
 We find that the fourth phase has a slightly larger standard deviation ($\sigma_4 = 1.14$) than the other phases ($\sigma_1=0.68$, $\sigma_2=0.68$, and $\sigma_3=0.65$) where $\sigma_{i}$ is the standard deviation for the $i$-th phase. 
 Jet breaks might be responsible for this large dispersion in the fourth phase. 
 In Figure 3a, we show the best fit model with the solid red lines.

\begin{figure*}[htbp]
\begin{center}
\begin{tabular}{c}
\begin{minipage}{0.50\hsize}
\begin{center}
\includegraphics[clip,width=9cm]{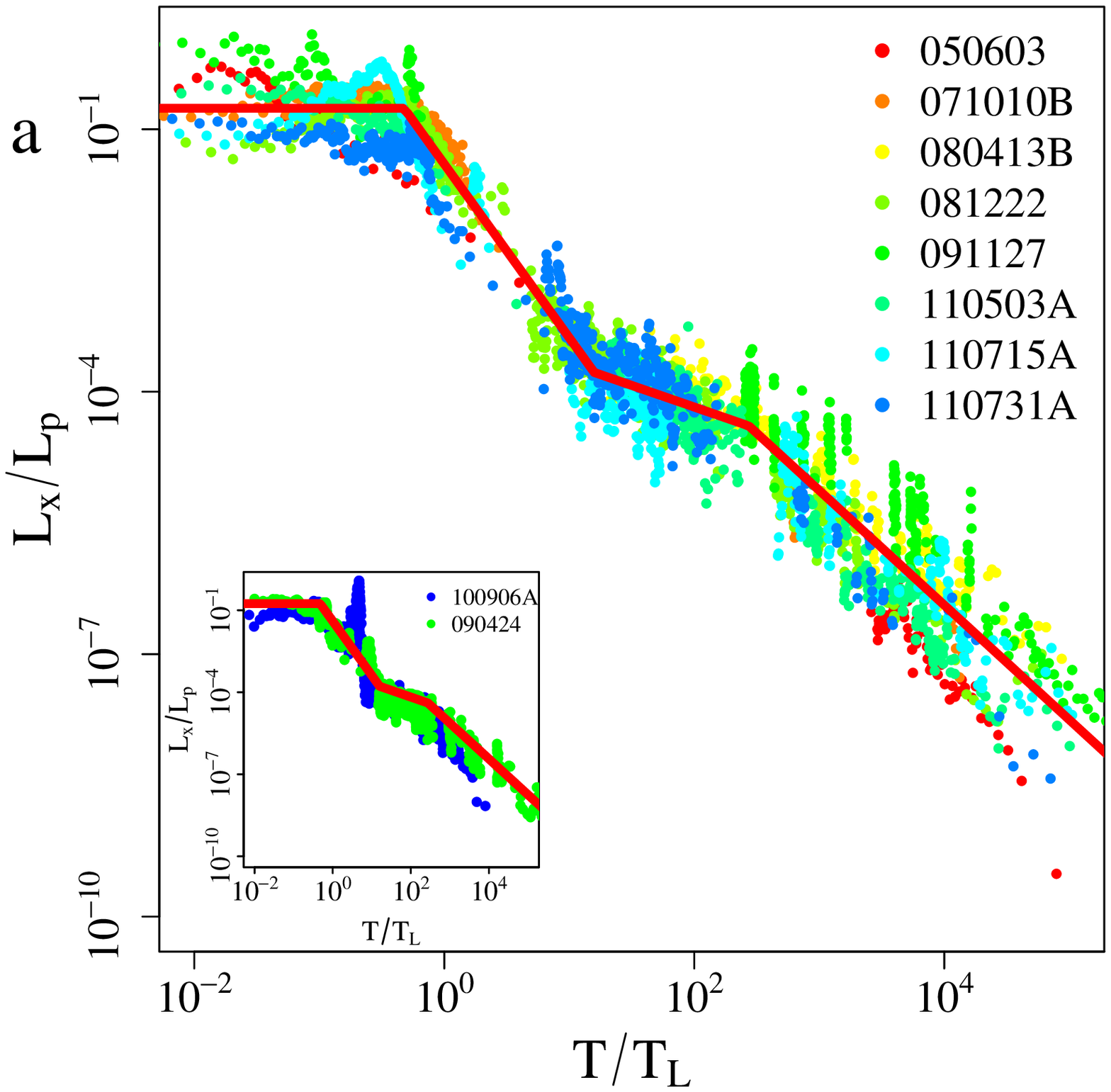}
\end{center}
\end{minipage}
\begin{minipage}{0.50\hsize}
\begin{center}
\includegraphics[clip,width=9cm]{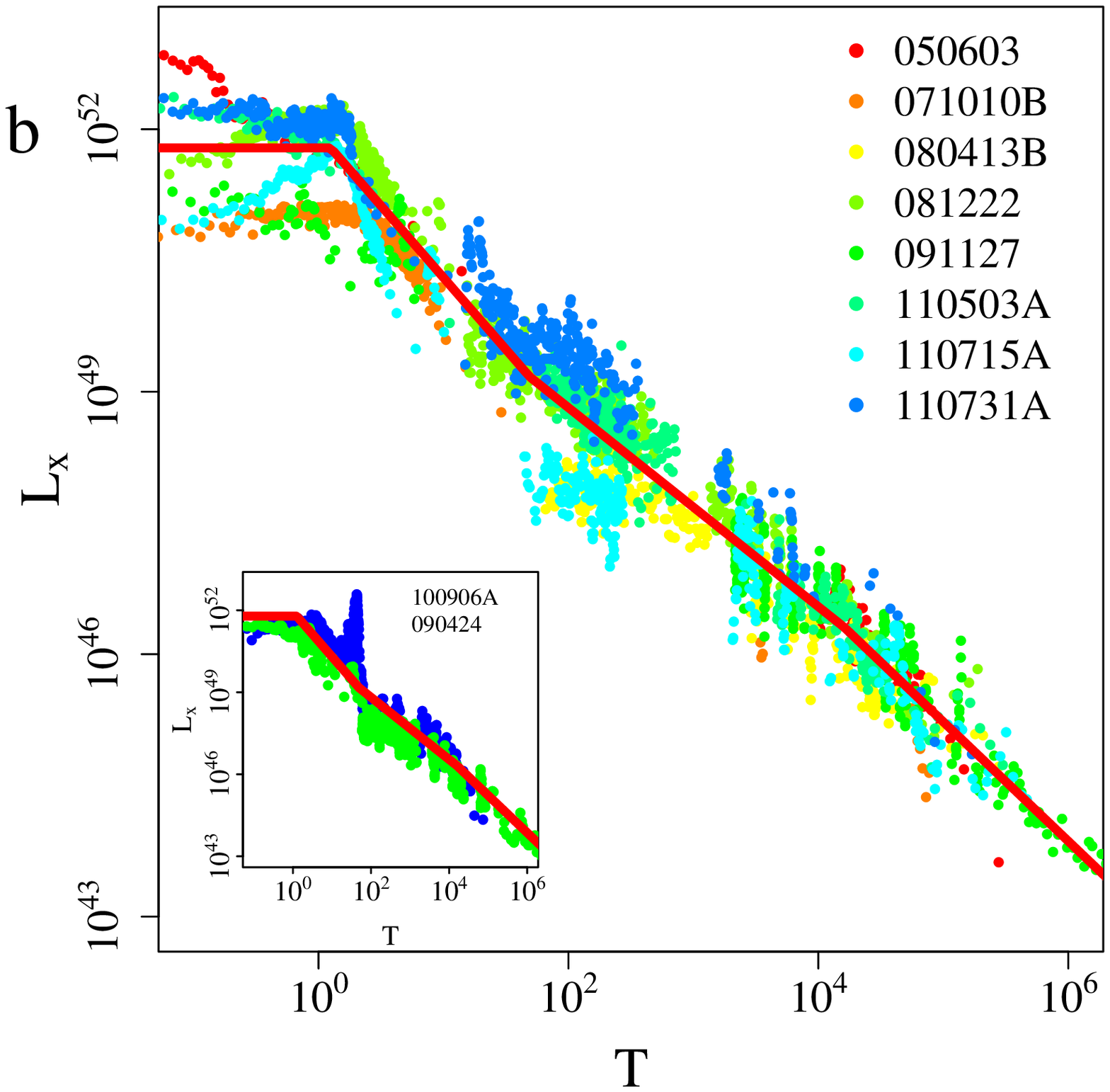}
 \end{center}
\end{minipage}
\end{tabular}
\end{center}
\caption{{\bf Normalized  (a) and unnormalized (b) BAT-XRT light curves for Type II long GRBs.}
Luminosities normalized with $L_{\rm p}$ as functions of the normalized time $T/T_{\rm L}$ are shown in the panel a. 
The panel b shows light curves constructed from the same data as in the panel a. 
Two events with strong X-ray flares are separately plotted in the inset. 
The best fit curve is indicated by the red solid line with three phases  described by power laws with different exponents (see text for details). 
 }
\label{TypeIIGRB}
\end{figure*}

To demonstrate how our scaling law works well,  we fit the same model to three light curve plots constructed from the same data with different normalizations, i.e., $F_{\rm X}(T_{\rm obs})$, $L_{\rm X}(T)$, and $L_{\rm X}(T_{\rm obs}/T_{90})/L_{\rm p}$. The standard deviation in each fitting in table 2 shows that the normalization in equation (1), i.e., $L_{\rm X}(T_{\rm obs}/T_{\rm L})/L_{\rm p}$ reduces the dispersion.
We summarize the standard deviation in each fitting in table 2.
As table 2 shows, 
the normalization in equation (\ref{eq}), $L_{\rm X}/L_{\rm p}$ versus $T/T_{\rm L}$, improves the fit of the model.  
For a visual comparison, we show  a relationship between the intrinsic luminosity $L_{\rm X}$ and time $T$ in Figure 3b.
Figures 3a and 3b show that our normalization really reduces the dispersion of light curves.

\begin{table}
\caption{Relationship between time scales, brightnesses and standard deviations 
for different normalizations. The normalization adopted in equation (\ref{eq}) is separately listed in the last row.
}
\begin{center}
\begin{tabular}{ccc}
\hline 
Time & Luminosity &  $\sigma$ \\
\hline \hline
$T_{\rm obs}$ & $F_{\rm X}$ & 0.93\\
$T$ & $L_{\rm X}$ & 1.02\\
$T/T_{90}$ & $L_{\rm X}/L_{\rm p}$ &1.01\\ \hline
$T/T_{\rm L}$ & $L_{\rm X}/L_{\rm p}$ & 0.84\\
\hline
\end{tabular}
\end{center}
\label{tab1}
\end{table}%

\section{Implication \& discussion}
Here we briefly discuss  a possible interpretation of the universal scaling law presented above by using the temporal evolution of photospheric emission of an optically thick jet propagating in the circumstellar matter (CSM). The dynamical evolution of such a jet relevant to observations is composed of four phases: the jet emergence phase, the stationary expansion phase, the blast wave phase, and the reverse shock phase. The last two phases are described by self-similar solutions\citep{Nakamura:2006,Shigeyama:2012}. In the first two phases, the central engine must be still active and the duration of the activity determines the time scale $T_{\rm L}$. The predicted temporal evolution of the luminosity in each phase is summarized as follows. The observed light curves in the first two phases seem to suggest the temporal evolution of the X-ray luminosity in the form of $L_{\rm X}\propto (1+T/T_{\rm L})^{-2}$. 
Furthermore, the exponents in the other phases can be deduced from the corresponding self-similar solutions for propagation of relativistic shocks in a stationary wind: $\alpha_3=1-\sqrt{3},\,\alpha_4<-3/2$. Detailed derivation is given in a separate paper (T.S. and R.T. in preparation). The rough agreement between  these exponents and those in equations (\ref{eq}) deduced from observations encourages such a modeling.

 Deeper understanding of the scaling law and refinement of the classification scheme will lead to more accurate determination of the intrinsic luminosities of LGRBs and then shed light on the more distant universe than ever reached by supernovae.

\section*{Acknowledgments}
 This work made use of data supplied by the UK Swift Science Data Centre at the University of Leicester.
 This work is supported in part by the Grant-in-Aid for Grant-in-Aid for Young Scientists (B) 
from the Japan Society for Promotion of Science (JSPS), No.24740116(RT).
%
%

\end{document}